\documentclass[twocolumn,amsmath,floatfix,amssymb,aps,prb,showpacs,superscriptaddress]{revtex4-1}
\usepackage{graphicx,natbib,multirow}
\usepackage{amsmath,tabularx}
\usepackage[colorlinks,bookmarks=false,citecolor=blue,urlcolor=blue]{hyperref}
\usepackage{color}

\begin{document}          

\title{Characterization of the soft X-ray spectrometer PEAXIS at BESSY II}

\author{Ch. Schulz}
\affiliation{Dept. Methods for Characterization of Transport Phenomena in Energy Materials, Helmholtz-Zentrum Berlin f\"{u}r Materialien und Energie, D-14109 Berlin, Germany}

\author{K. Lieutenant}
\affiliation{Dept. Methods for Characterization of Transport Phenomena in Energy Materials, Helmholtz-Zentrum Berlin f\"{u}r Materialien und Energie, D-14109 Berlin, Germany}

\author{J. Xiao}
\affiliation{Dept. Highly Sensitive X-Ray Spectroscopy,\\Helmholtz-Zentrum Berlin f\"{u}r Materialien und Energie, D-14109 Berlin, Germany}

\author{T. Hofmann}
\affiliation{Dept. Methods for Characterization of Transport Phenomena in Energy Materials, Helmholtz-Zentrum Berlin f\"{u}r Materialien und Energie, D-14109 Berlin, Germany}

\author{D. Wong}
\affiliation{Dept. Methods for Characterization of Transport Phenomena in Energy Materials, Helmholtz-Zentrum Berlin f\"{u}r Materialien und Energie, D-14109 Berlin, Germany}

\author{K. Habicht}
\affiliation{Dept. Methods for Characterization of Transport Phenomena in Energy Materials, Helmholtz-Zentrum Berlin f\"{u}r Materialien und Energie, D-14109 Berlin, Germany}
\affiliation{Institut f\"{u}r Physik und Astronomie, Universit\"{a}t Potsdam, D-14476 Potsdam, Germany}
\email[Corresponding author: ]{habicht@helmholtz-berlin.de}

\date{\today}

\begin{abstract}
The performance of the recently commissioned spectrometer PEAXIS for resonant inelastic soft X-ray scattering (RIXS) and X-ray photoelectron spectroscopy (XPS) and its hosting beamline U41-PEAXIS at the BESSY II synchrotron are characterized. The beamline provides linearly polarized light from 180\,eV\,-\,1600\,eV allowing for RIXS measurements in the range of 200\,eV\,-\,1200\,eV. The monochromator optics can be operated in different configurations for the benefit of either high flux, providing up to $10^{12}$\,photons/s within the focal spot at the sample, or high energy resolution with a full width at half maximum of $<$\,40\,meV at an incident photon energy of $\sim$400\,eV. This measured total energy resolution of the RIXS spectrometer is in very good agreement with the theoretically predicted values by ray-tracing simulations. PEAXIS features a 5 m long RIXS spectrometer arm that can be continuously rotated about the sample position by $106^\circ$ within the horizontal photon scattering plane, thus enabling the study of momentum-transfer-dependent excitations. To demonstrate the instrument capabilities, d-d excitations and magnetic excitations have been measured on single-crystalline NiO. Measurements employing a fluid cell demonstrate the vibrational progression in liquid acetone. Planned upgrades of the beamline and the RIXS spectrometer that will further increase the energy resolution by $20-30\%$ to $\sim100$\,meV at 1000\,eV incident photon energy are discussed.
\end{abstract}

\maketitle    

\section{Introduction}

The electronic structure and dynamics determine important fundamental properties of materials such as charge and spin transport, novel phases of matter and emergent phenomena originating in strong electronic correlations. Thus probing electronic excitations and the coupling of electrons to lattice, orbital and spin degrees of freedom provides an invaluable tool to understand the microscopic mechanisms which determine macroscopic functionalities for technological applications.

Resonant inelastic X-ray scattering (RIXS) at brilliant third-generation synchrotron radiation sources nowadays is a mature technique known to access element-specific information on the electronic structure and dynamics as well as to gain momentum-resolved information on quasiparticle excitations which couple to the electronic states. Being a photon-in, photon-out spectroscopic technique, RIXS is bulk-sensitive and thus well suited to investigate ordered solid state materials. In the soft X-ray regime, RIXS allows for the resonant excitations of electronic states in important elements such as C, N, O and the series of 3d transition metal elements present in a broad range of materials of fundamental and technological interest.
The technique has recently been applied to address important scientific questions in quantum materials with break-through results identifying charge collective modes as acoustic plasmons in electron-doped copper oxide superconductors \citep{Hepting2018}, probing multi-spinon excitations in spin 1/2 one-dimensional Heisenberg antiferromagnets \citep{Schlappa2018} or extracting the electron-phonon coupling strength in oxide heterostructures \citep{Meyers2018}. Beyond numerous applications for RIXS in solid state physics, intra- and inter-molecular couplings can be studied in molecular systems either in the liquid phase \citep{VazdaCruz2019} or in the gas phase \citep{Couto2017}.

Numerous soft X-ray RIXS spectrometers have been built at synchrotron facilities in the recent past, such as SAXES at SLS, PSI \citep{Strocov2010}, SEXTANTS at SOLEIL \citep{Chiuzbaian2014}, RIXS at BL05A at the TLS \citep{Huang2018}, iRIXS at ALS \citep{Chuang2017}, SIXS at NSLS II \citep{Jarrige2018} and ID32 at ESRF \citep{Brookes2018}, or are currently either in commissioning or in the planning stage, such as I21 at Diamond Light Source \citep{I21Diamond}, VERITAS at MAX IV \citep{VeritasMAXIV} and METRIX at BESSY II \citep{Pietzsch2018} to address the increasing demand for RIXS capabilities.

The new spectrometer PEAXIS (Photo Electron Analysis and resonant X-ray Inelastic Spectroscopy) at the third generation synchrotron BESSY II adds to the capacity of RIXS spectroscopy for the international user community. It has been built with a particular focus on solid state applications and allows for spectroscopy of liquids encapsulated in sealed cells. In addition, X-ray Photoelectron Spectroscopy (XPS) capabilities are offered. In this article, important instrumental parameters of the hosting beamline and the spectrometer are presented and the overall performance of PEAXIS is demonstrated on selected experimental examples.

\section{Instrument layout}

\subsection{Beamline design}

\begin{figure*}[ht]
\centering
\includegraphics[width=11cm]{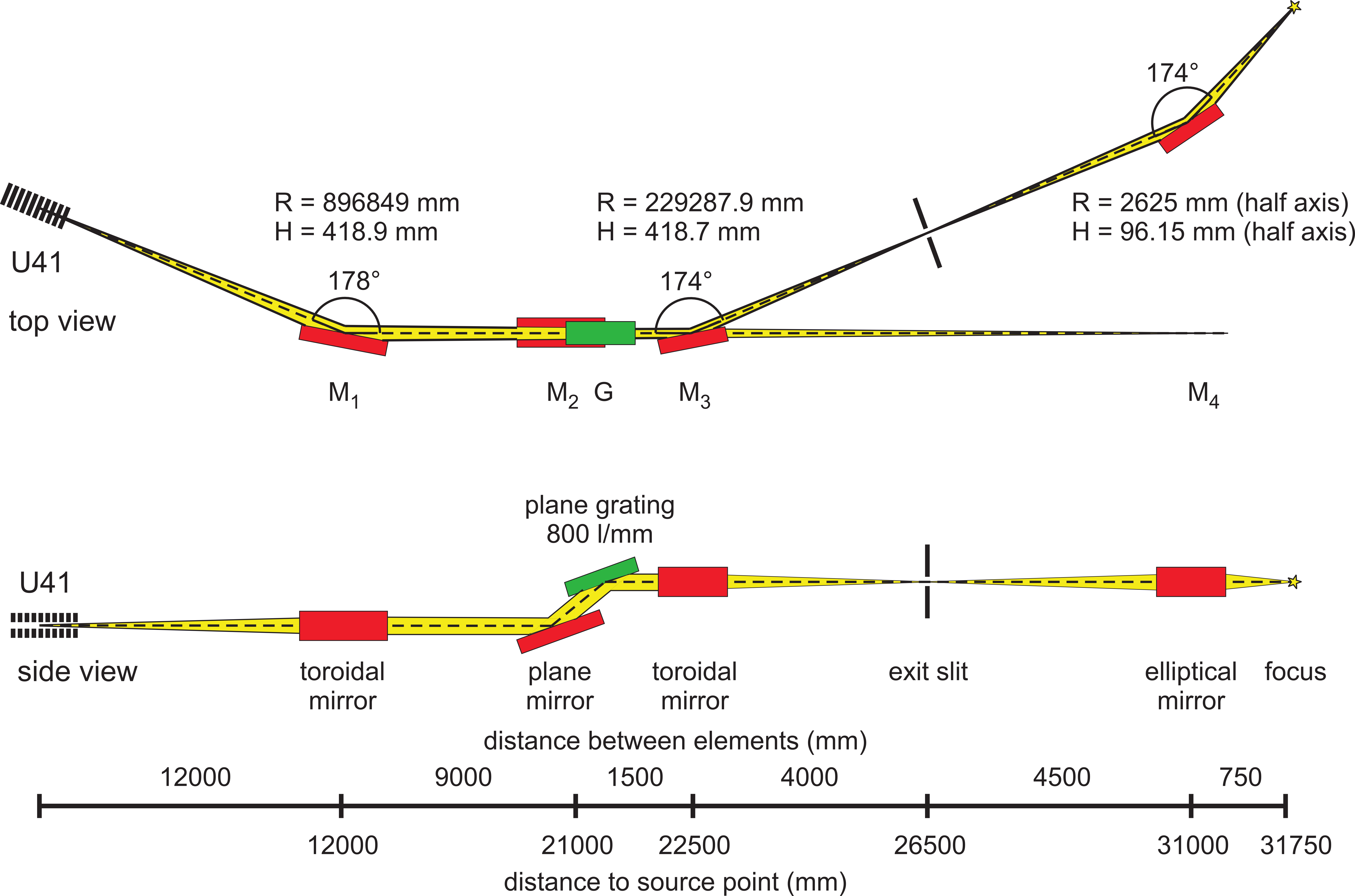}
\caption{Layout of the U41-PEAXIS beamline at BESSY II \citep{BrzhezinskayaReport}.}
\label{fig:fig1}
\end{figure*}

PEAXIS is a fixed station installed at undulator U41 of BESSY II which has 41\,mm period length and a minimum gap of 15.5\,mm and delivers linearly polarized light with horizontal polarization. The optical layout of the beamline is shown in Fig.\,\ref{fig:fig1}. It is based on a plane grating monochromator (PGM) design, which can achieve high energy resolution, high flux and a wide energy range coverage by employing a single grating only. A first toroidal mirror ($M1$) collimates the divergent beam from the undulator source in the horizontal plane and provides a parallel beam in the vertical plane. The next optical element is the monochromator which disperses the beam according to its photon energies. It consists of a plane pre-mirror ($M2$) and a plane blazed grating ($G$) assembly in which the grating has a groove density of 800 lines/mm. The monochromator can be aligned to different fixed-focus constants $c_{\rm{ff}}=\cos\beta/\cos\alpha$. Here $\alpha$ and $\beta$ are the incident and exit angles at the monochromator grating (cf. Section \ref{sec:RIXS}). A second toroidal mirror ($M3$) then focuses the parallel but spatially extended beam in both the vertical and the horizontal planes onto the exit slit, providing monochromatic light after the slit. Finally, an elliptical mirror ($M4$) refocuses the beam onto the sample position in the PEAXIS sample chamber. \\

\subsection{Photon flux, focal spot size and energy resolution}

By combining different $c_{\rm{ff}}$ values with appropriate exit slit widths, the beamline can be tuned for either higher energy resolution or increased photon flux. The three typically chosen modes of operation are a high-flux mode ($c_{\rm{ff}}=2.25$, slit size 20\,$\mu$m), a standard mode ($c_{\rm{ff}}=3$, slit size 10\,$\mu$m) and a high-resolution mode ($c_{\rm{ff}}=5$, slit size 5\,$\mu$m), though intermediate combinations of $c_{\rm{ff}}$ and slit size can be freely chosen to match the experimental requirements. For the three typical settings, the incident photon flux at the focal point of the beamline at the sample position within the PEAXIS chamber have been determined by total electron yield measurements on a clean Au sample and by recording the current on a windowless GaAsP photodiode (Hamamatsu G1127-04). These measurements were carried out for different undulator harmonics covering the whole energy range of the beamline.\\

\begin{figure}[h]
\includegraphics[width=8cm]{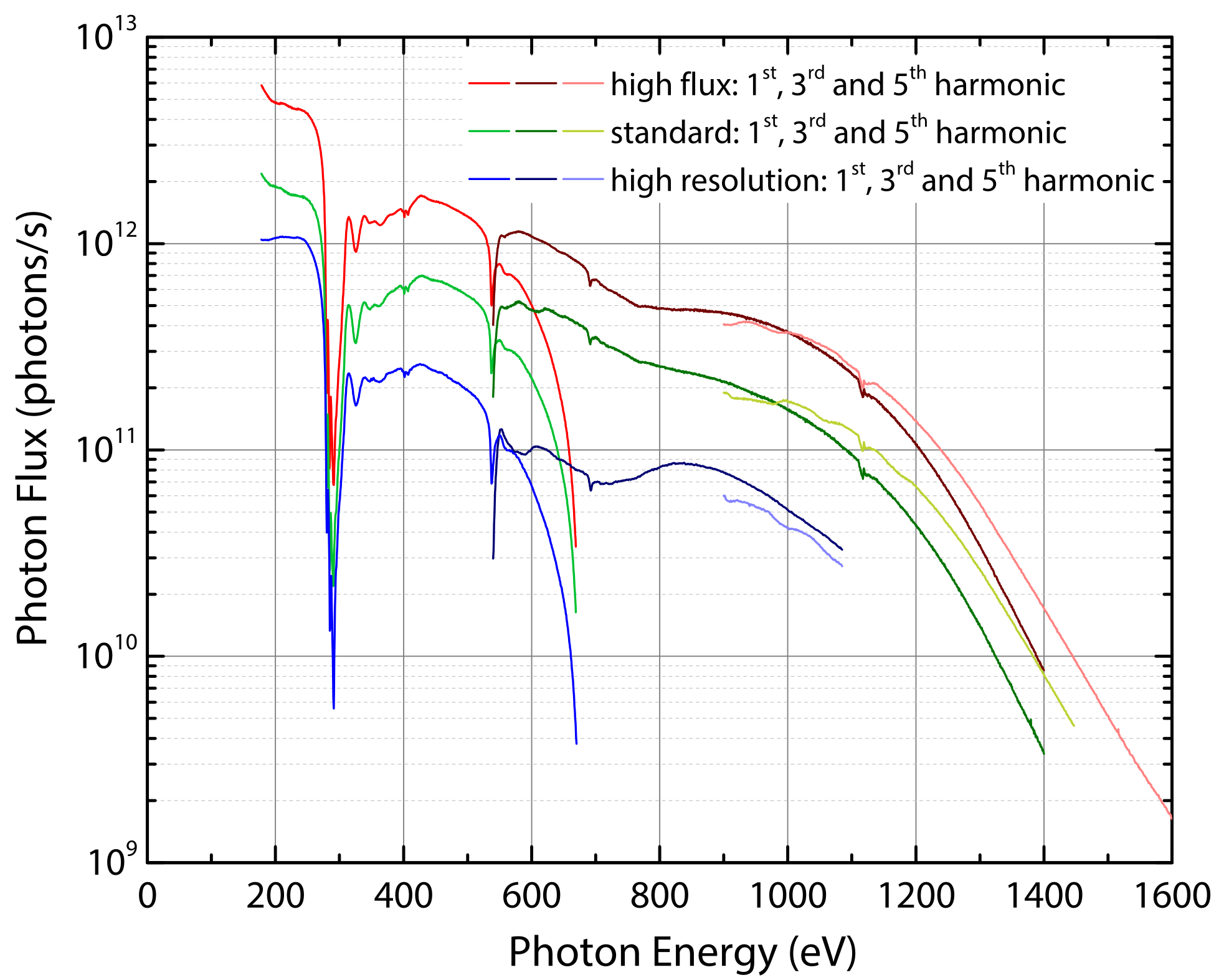}
\caption{X-ray photon flux at the sample position for the three operation modes of the beamline and 1st, 3rd and 5th harmonics from the undulator. The measurements were taken under normal multi-bunch operation of the BESSY II storage ring with a current of 250\,mA and an opening of the beamline entrance slits of 2\,mm\,x\,3\,mm (standard use).}
\label{fig:fig2}
\end{figure}

The results of these flux measurements are shown in Fig.\,\ref{fig:fig2} for the three typical modes of operation. Note that for the high-resolution mode ($c_{\rm{ff}}=5$ and slit opening 5\,$\mu$m), the photon energy of the beamline is limited to 1085\,eV, as a significant fraction of the photon beam will not be reflected by the pre-mirror ($M2$ in Fig.\,\ref{fig:fig1}) in front of the grating for higher incident photon energies.

\begin{figure}[h]
\centering
\includegraphics[width=8cm]{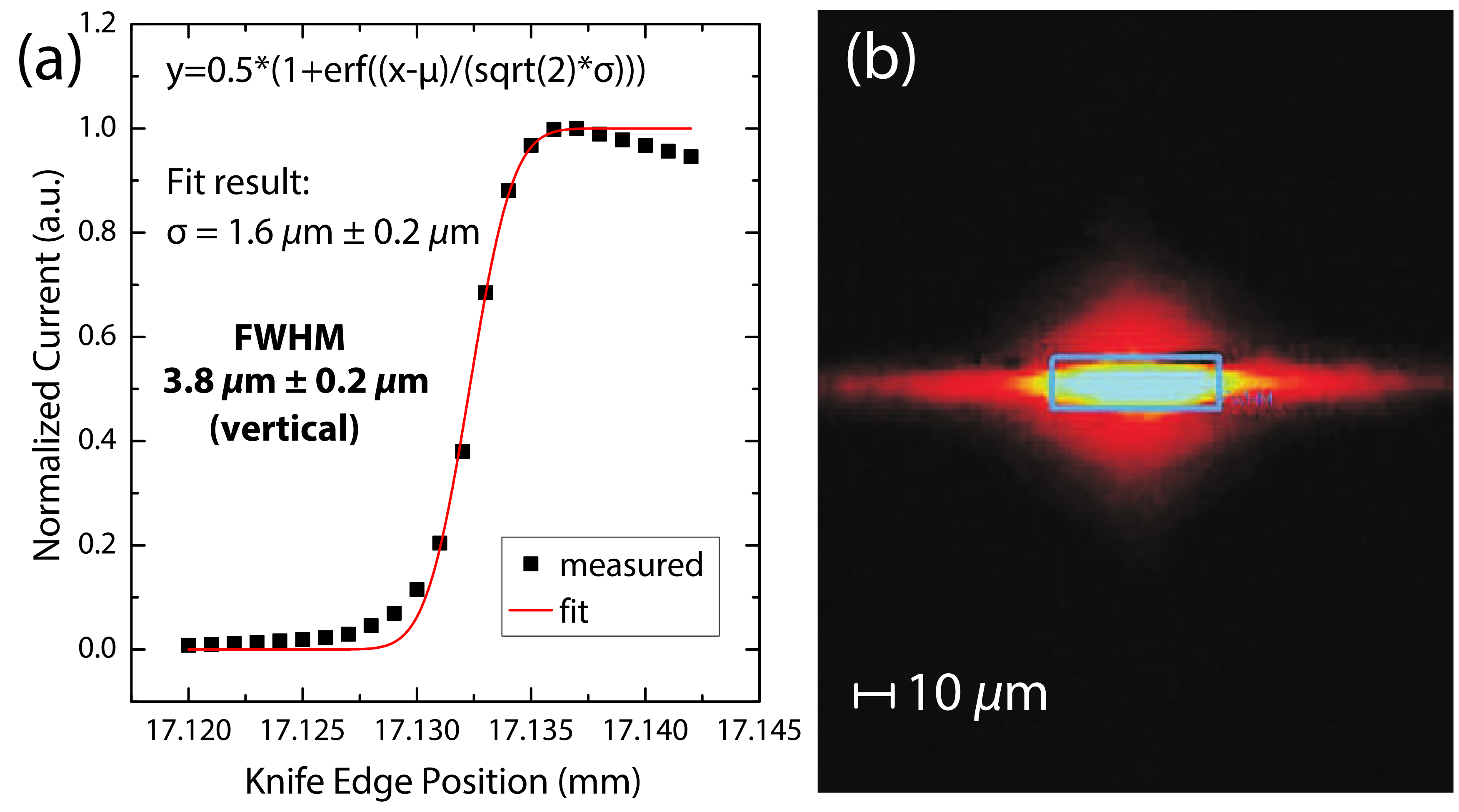}
\caption{(a) Fit of knife-edge measurement data to determine the (vertical) Gaussian FWHM of the focal spot size. (b) Two-dimensional image of the focal spot of the beamline at the sample position, recorded by a CCD with high spatial resolution in the high-flux mode of operation of the beamline.}
\label{fig:fig3}
\end{figure}

The focal spot size was determined by measuring the induced electric current on a gold coated knife-edge moved incrementally into the beam until reaching current saturation for full coverage of the focal spot in both vertical and horizontal directions. The resulting step functions were fitted to obtain the full-width half-maximum (FWHM) of the assumed Gaussian spot profiles (shown for the vertical spot size in Fig.\,\ref{fig:fig3}(a)). For the high-flux mode of operation, we thus obtain a maximum focal spot size of 3.8\,$\mu$m x 12.4\,$\mu$m FWHM (vertical x horizontal). This small vertical spot size is important for a good energy resolution of the RIXS spectrometer, because it defines the deviation from the optimal incident angle $\alpha$ to the grating in the RIXS spectrometer.\\

Additionally, a two-dimensional image of the focal spot was recorded by using a dedicated vacuum chamber placed at the sample position prior to the installation of the PEAXIS sample chamber. This measurement setup allows to position a scintillator detector which converts the X-ray image into visible light at the focal spot of the beamline. The image is subsequently magnified by an optical zoom lens and detected by a CCD camera, both being located outside the vacuum chamber. Compared to the RIXS setup, for example, the lens in this setup can be positioned very close to the scintillator to obtain a numerical aperture $n\sin\phi$, with refractive index $n$ and aperture angle $\phi$, which provides the required high spatial resolution \citep{Martin2006}. Nevertheless, there is a discrepancy between the focal spot size determined by this method compared to that determined from the knife-edge measurements which we attribute to inherent scattering of the visible light within the scintillator and the glass window behind it.

The resolving power of the beamline was determined from absorption measurements in N$_2$ and Ne gas, the latter of which is also used for the calibration of the monochromator setup for an absolute value of the incident photon energy. These two gases have resonance energies of $\sim400$\,eV and $\sim870$\,eV, respectively, which allows to characterize the beamline performance at two pertinent incident photon energy regions in the PEAXIS energy range of operation.

\begin{figure}[b]
\centering
\includegraphics[width=8cm]{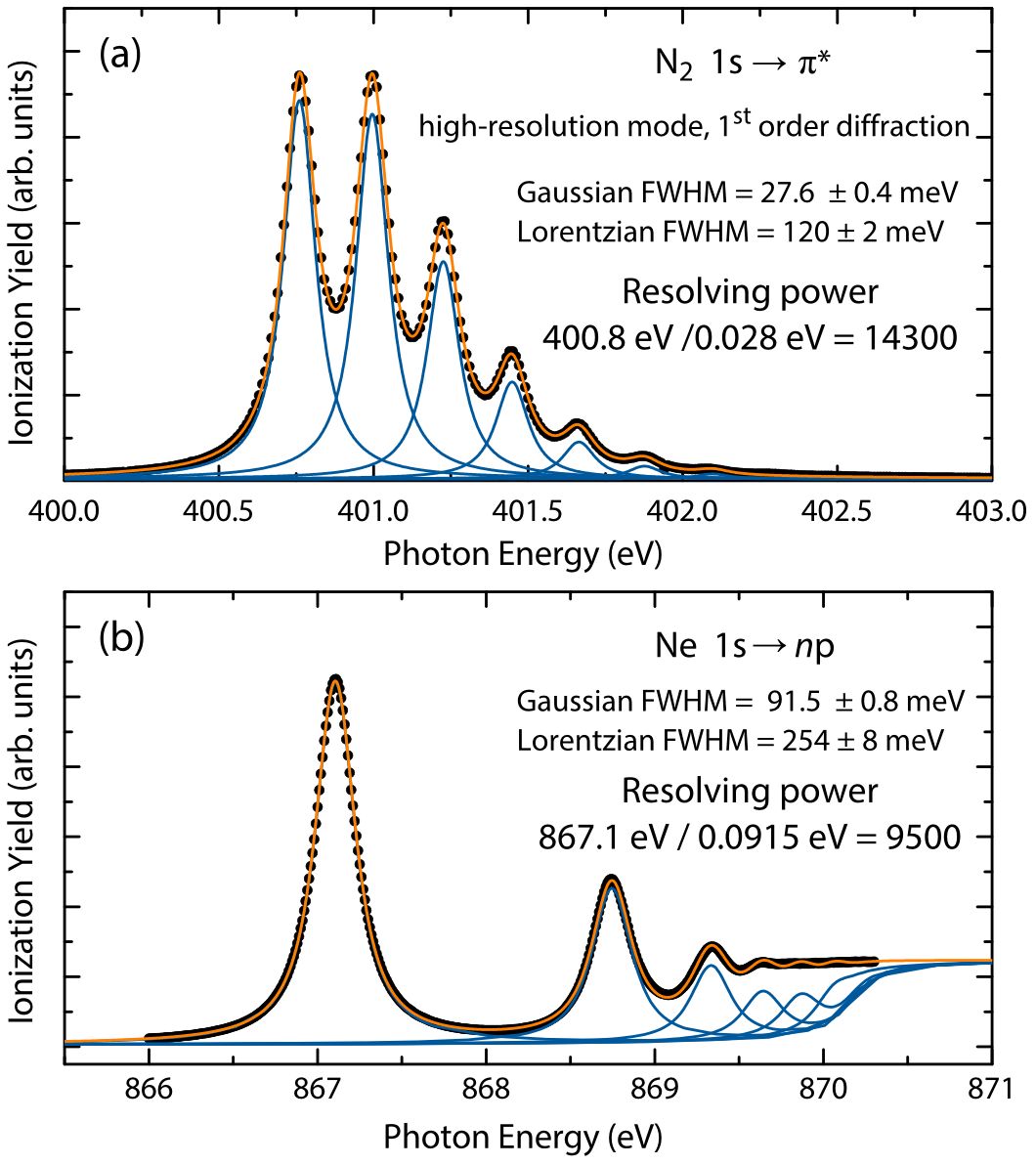}
\caption{Beamline energy resolution as measured from (a) N$_2$ and (b) Ne gas photoabsorption. Both spectra were measured in first diffraction order and in the high-resolution mode of the beamline. For details, please see text.}
\label{fig:fig4}
\end{figure}

\begin{table*}
\centering
\resizebox{15cm}{!}{
\begin{tabular}{c|c|c|c|c|c}
 operation    &\multirow{2}{*}{c$_{\rm{ff}}$}        & slit size       & $\Delta E$ @\,400.8 eV &\multirow{2}{*}{ $E/\Delta E$ @\,400.8 eV} & photon flux\,@\,400.8\,eV \\
 mode& & ($\mu$m)& (meV) & & (s$^{-1}$\,x\,0.1$\%$\,BW\,x\,100\,mA)\\
\hline
high flux & 2.25&20&61.2 $\pm$ 0.5& 6500 & 3.51 x 10$^{12}$\\
standard & 3 & 10 & 37.7 $\pm$ 0.3 & 10500 & 2.42 x 10$^{12}$\\
high resolution & 5&5&27.6 $\pm$ 0.4&14300& 1.33 x 10$^{12}$\\
\hline\hline
 operation  &\multirow{2}{*}{c$_{\rm{ff}}$}   & slit size & $\Delta E$ @\,867.1 eV &\multirow{2}{*}{ $E/\Delta E$ @\,867.1 eV} & photon flux  @\,867.1 eV     \\
mode & & ($\mu$m) & (meV) & & (s$^{-1}$ x 0.1$\%$\,BW x 100\,mA)\\
   \hline
high flux & 2.25&20&156 $\pm$ 2& 5500 &1.05 x 10$^{12}$\\
standard & 3 & 10 &106 $\pm$ 2 & 8200 &7.49 x 10$^{11}$\\
high resolution & 5&5&91.5 $\pm$ 1&9500&3.16 x 10$^{11}$\\
\end{tabular}}
\caption{Energy resolution, resolving power and photon flux of the beamline for the high flux, standard and high resolution modes of operation, measured on N$_2$ and Ne gas absorption lines for energy regions around 401\,eV and 867\,eV, respectively.}
\label{tab:table1}
\end{table*}

For the lower energy region, the photon absorption of N$_2$ at the 1s$\rightarrow\pi^*$ resonance around 401\,eV leads to a fine structure due to vibrational levels of the $\pi^*$ state, whose peak positions and linewidths are routinely used to determine the intrinsic instrumental resolution of soft X-ray beamlines \citep{Feifel2004}. The recorded photoabsorption spectrum of N$_2$ for a gas cell pressure of 2 x 10$^{-3}$ hPa and the high-resolution mode of the PEAXIS beamline is shown in Fig. \ref{fig:fig4}(a). The spectrum is characterized by the appearance of eight vibrational levels which are fitted each with a convolution of a Gaussian lineshape representing the intrinsic energy resolution of the beamline (in FWHM) and a Lorentzian lineshape to represent the Nitrogen 1s core hole lifetime broadening. Measurements with different settings for $c_{\rm{ff}}$, slit size and diffraction order of the undulator have been used to determine the Lorentzian width to a mean value of 120$\pm$\,2\,meV FWHM, which is in good agreement with values found in literature \citep{Feifel2004,Kato2007N2}. Based on this value for the core hole lifetime broadening in N$_2$, the Gaussian FWHM of the beamline was found to be 27.6$\pm$0.4\,meV. For the incident energy of 400.8\,eV, this energy resolution of $\sim$\,28\,meV thus corresponds to a resolving power $E/\Delta E$ exceeding 14300, which is close to the resolution limit of this method due to the uncertainty of the Lorentzian broadening. To circumvent this limitation, the combined energy resolution of the beamline and the RIXS spectrometer are evaluated by linewidth measurements of the elastically scattered X-rays on a polished NiO crystal surface (see Section \ref{sec:RIXS}).

At higher incident photon energy, the energy resolution of the beamline was determined from the 1s$^{-1}\rightarrow$ \textit{n}p Rydberg series of Ne under the same experimental conditions as for the N$_2$ measurements discussed above. Here, a Lorentzian FWHM of 254\,$\pm$8\,meV was obtained by fitting the 1s$^{-1}\rightarrow$ 3p peak for different beamline settings. This Lorentzian width is in good agreement with values found in literature \citep{Kato2007Ne}. The Ne photoabsorption spectrum shown in Fig. \ref{fig:fig4}(b) was then fitted with a series of Voigt profiles for the 1s$^{-1}\rightarrow$ \textit{n}p (\textit{n}=3-8) Rydberg series on top of a nonlinear background at high energies related to the Ne 1s core ionization threshold at 870.16\,eV. A common Gaussian FWHM of 91.5$\pm$0.8\,meV was so determined as the intrinsic energy resolution at the incident energy of 867.1\,eV, corresponding to a resolving power of 9500 for the high-resolution mode of operation. The obtained resolution parameters for the typical operation modes of the beamline from N$_2$ and Ne measurements are listed in Table \ref{tab:table1}.

\section{PEAXIS instrument}

\begin{figure*}[t!]
\centering
\includegraphics[width=15cm]{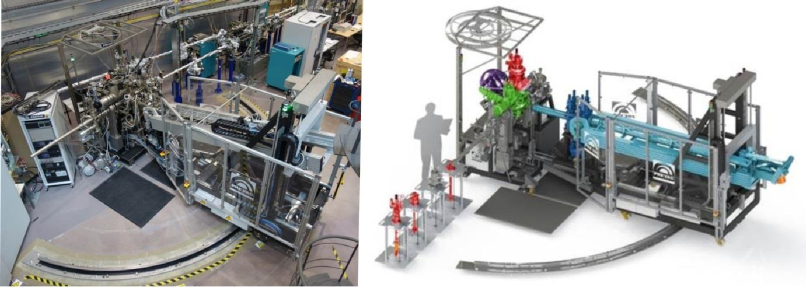}
\caption{(Left) Photograph of the PEAXIS station installed and in operation in the BESSY II experimental hall. (Right) Schematic diagram of PEAXIS highlighting the main components: The rotatable arm of the RIXS spectrometer (light blue) together with the grating chamber (dark blue), the RIXS detector (orange), the sample chamber (green) with the available sample manipulators (red), the XPS detector (violet) and the load lock and sample preparation chamber (left side in front of the sample chamber).}
\label{fig:fig5}
\end{figure*}

PEAXIS combines two important types of measurement techniques, namely wave vector-dependent Resonant Inelastic X-ray Scattering (RIXS) and X-ray Photoelectron Spectroscopy (XPS) in a single instrument. The RIXS instrument is based on an in-house design concept previously discussed \citep{LieutenantJElSpec}, while the XPS analyzer is a commercial "PHOIBOS 150 EP" hemispherical analyzer from SPECS \citep{SPECS}. The PEAXIS UHV components, mechanics and electronics were commercially built by PREVAC \citep{Prevac}. A photograph of the PEAXIS spectrometer as installed in the BESSY II experimental hall is shown on the left side of Fig. \ref{fig:fig5}, with the main components highlighted in color in the schematic drawing on the right side of the figure.
 
An important feature of PEAXIS is the possibility to rotate the RIXS spectrometer arm \textit{continuously} around the sample for $2\theta$ scattering angles between 33$^\circ$ - 139$^\circ$, providing a wave vector coverage of 0.058\,\AA$^{-1}$ - 0.190\,\AA$^{-1}$ at 200\,eV and 0.345\,\AA$^{-1}$ - 1.139\,\AA$^{-1}$ at 1200\,eV incident photon energy. The RIXS spectrometer operates with two spherical variable line space (VLS) gratings for energy resolving measurements in the energy range of 200\,eV - 600\,eV and 400\,eV - 1200\,eV, respectively. The complete RIXS design as well an analytical calculation of the required parameters for the line densities of the VLS gratings and a numerical optimization of the overall parameters such as the total length of spectrometer arm and considerations on optimum distances between sample position and analyzer grating at a given photon energy have been described previously \citep{LieutenantJElSpec,LieutenantJPhys2016}. A more detailed description of the main components of PEAXIS is given in the following subsections \ref{sec:SampleChamber} - \ref{sec:XPS}.

\subsection{Sample chamber and sample environments}\label{sec:SampleChamber}

The sample chamber centered at the focal point of the beamline is preceeded by a load-lock chamber that is separately pumped to achieve ultra-high vacuum (UHV) conditions prior to the transfer of the sample to the measurement position inside the sample chamber. The load-lock chamber hosts a six-fold sample holder that accommodates standard flag-style sample holders of 18\,mm\,x\,21\,mm size typically made out of Tantalum or Copper onto which the samples of optimal thickness 1.5\,mm are attached by means of Cu tape or conductive Ag paint. Three out of the six positions on the load-lock sample holder can optionally be heated up to 800\,K, which may be used for thermal annealing purposes prior to the actual measurements. A mechanical cleaving tool allows for \textit{in-situ} cleaving of the samples at a pressure $<5$x10$^{-9}$\,mbar if uncontaminated sample surfaces are required. Alternatively, the load-lock is equipped with an Argon ion sputtering tool that can be used for a sputtering treatment of the sample surfaces.

\begin{figure}[h!]
\centering
\includegraphics[width=6.5cm]{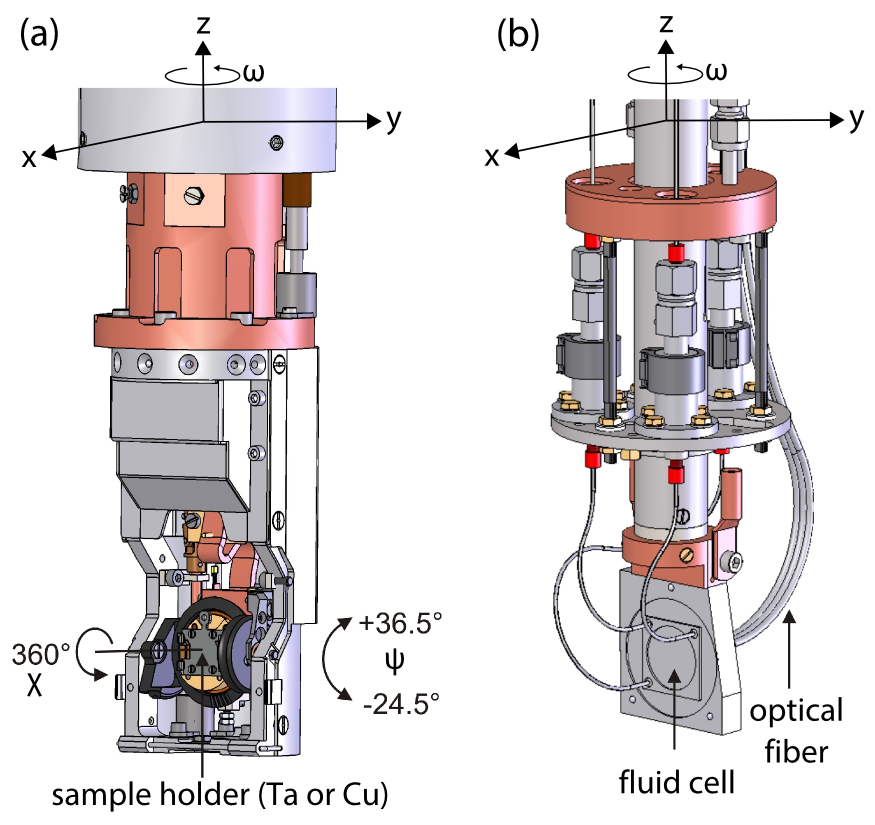}
\caption{PEAXIS sample manipulators and motion degrees of freedom: (a) Solid sample manipulator, here shown for the low temperature range (CCR) and (b) liquid sample manipulator.}
\label{fig:fig6}
\end{figure}

A base stage located at the top of the sample chamber allows for the sample positioning by four degrees of freedom, i.e. three orthogonal translations along the $x$ (along the photon beam), $y$ and $z$ directions and a rotation about an axis parallel to the $z$-direction ($\omega$) as shown in Fig. \ref{fig:fig6}.

Three different sample manipulators can be inserted into this base stage for different kinds of samples (solid or liquid) and provide different sample temperature ranges. A closed-cycle helium (CCR) manipulator covers a temperature range from 10\,K\,-\,300\,K, while a second manipulator is available for the high temperature range from 77\,K up to 1000\,K. For both manipulators, sample temperatures below room temperature can be controlled with a precision of 0.1\,K. For temperatures above 300\,K, the sample temperature is derived from the heating power, which currently restricts the precision to $\pm$5\,K below 350\,K and above 700\,K and to $\pm$10\,K in the intermediate temperature range. An upgrade to a direct sample temperature measurement with higher precision over the whole temperature range 300\,K\,-\,1000\,K is planned for the near future. The two manipulators for solid samples provide two additional rotations (labeled $\chi$ and $\psi$ in Fig. \ref{fig:fig6}(a)) around two orthogonal axes in the $x$-$y$ plane defined by the base stage. Hence, solid samples can be positioned by a total of six degrees of freedom. The third manipulator shown in Fig. \ref{fig:fig6}(b) is designed for the measurement of liquids that are sealed in a fluid micro-cell with a thin membrane (about 100\,nm thick) that allows for photon transmission. This liquid sample manipulator does not have any additional tilting/azimuthal rotation capabilities and allows to align the cell with the four degrees of freedom of the base stage.

\subsection{RIXS spectrometer}\label{sec:RIXS}

\begin{figure}
\centering
\includegraphics[width=\columnwidth]{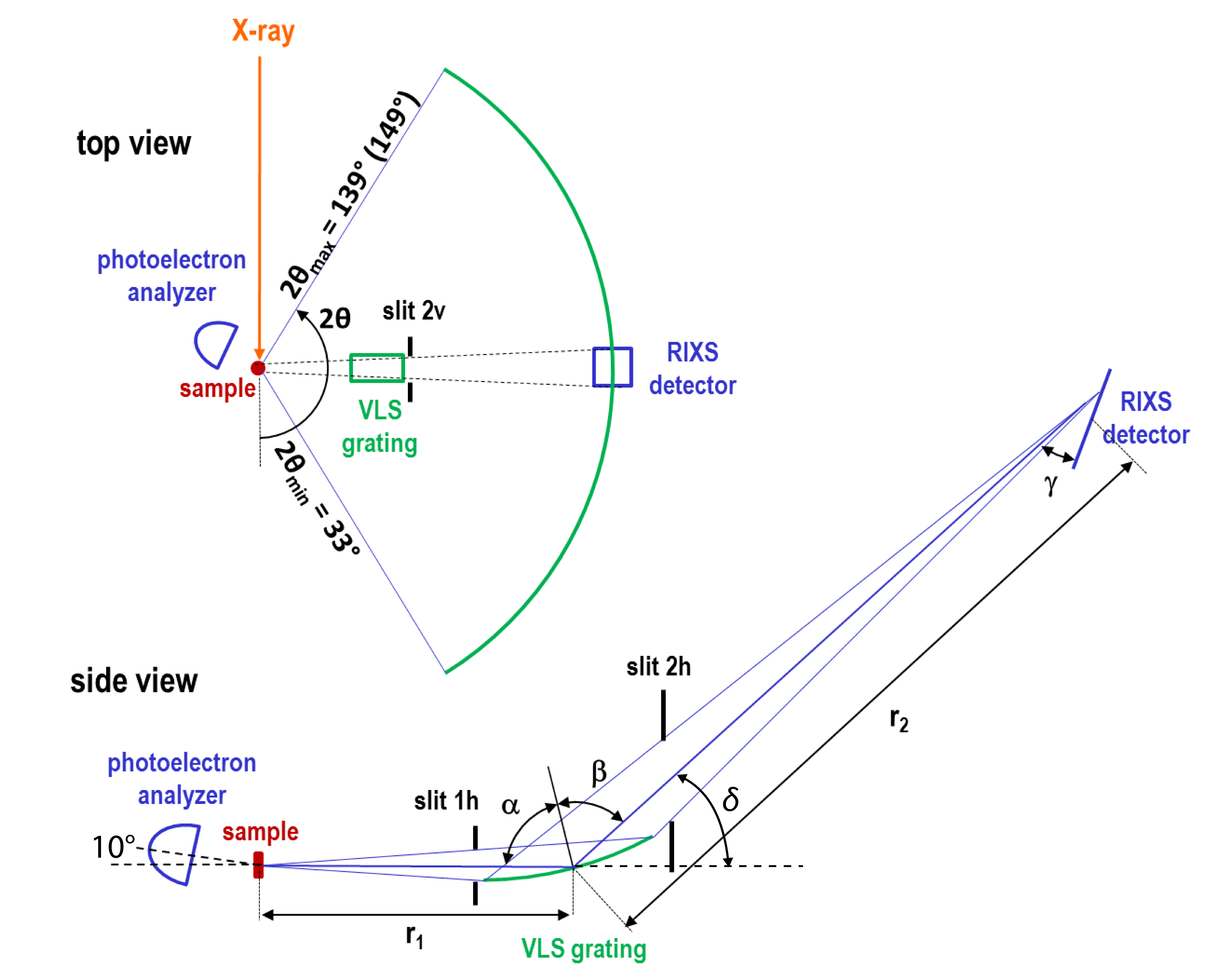}
\caption{Schematic drawing of the RIXS spectrometer setup in the horizontal (top view, top panel) and in the vertical (side view, bottom panel) photon scattering planes, with angles and distances as discussed in the text.}
\label{fig:fig7}
\end{figure}

The RIXS spectrometer incorporates a 5\,m long arm which can be rotated continuously about the sample position without breaking vacuum. A schematic drawing of the setup is shown in Fig. \ref{fig:fig7}. The rotation or scattering angle $2\theta$ defined as the angle between the incident photon beam and the emitted photons from the sample in the direction of the analyzer and detector can be varied between 33$^\circ$ and 139$^\circ$ with a precision of better than 0.03$^\circ$. This angular range corresponds to accessible wave vectors $Q$ from 0.058\,\AA$^{-1}$ - 0.190\,\AA$^{-1}$ and 0.345\,\AA$^{-1}$ - 1.139\,\AA$^{-1}$ for photon energies of 200\,eV and 1200\,eV, respectively. The angular range can be extended to 149$^\circ$, provided the distance $r_1$ between the sample and the grating used to analyze the emitted photon energy is larger than 1071\,mm, which presently requires to use incident photon energies between 550\,-\,600\,eV and the low energy grating.

The RIXS spectrometer is equipped with two spherically shaped VLS gratings. Both of them are mounted in the grating chamber on a grating manipulator. This manipulator has five degrees of freedom that allow to position the selected grating based on the calculated values for the chosen photon energy. The angle $\alpha$ between the photon beam emitted from the sample and the normal to the grating surface can be varied between 84$^\circ$ - 90$^\circ$ with a precision $\leq10^{-5}$ degrees. A second rotation around the direction of the incident photon beam positions the grating such that its spherically shaped surface is symmetric with respect to the central photon beam. Three translations are necessary to align the grating with respect to height and distance to the sample and to switch between the two VLS gratings mounted on the grating holder. The grating chamber (together with the grating manipulator) can be moved along the RIXS arm. By that movement, the distance between the sample and the center of the grating ($r_1$) can be adjusted from 650\,mm to 1250\,mm with an accuracy of 0.01\,mm. To prevent any absorption of the soft X-rays in air, the area between the sample and the grating chambers is kept under vacuum within a bellow with a differential pumping chamber.\\
Both gratings are mounted with their gold coated active surfaces facing upwards and thus the photons are dispersed in the vertical direction and detected at the distance $r_2$ by a cooled CCD detector (model iKon-L, cooled to -100$^\circ$ by a five-stage Peltier cooler to minimize the detector dark current to realize long exposure times) from Andor \citep{Andor}. Similar to $r_1$, the distance $r_2$ between the grating and the detector can be adjusted to within 0.01\,mm. The available distance ranges between 2012\,mm and 4080\,mm. with a tolerance of 0.01 mm. The angle $\delta$ (see Fig. \ref{fig:fig7}) between the horizontal plane and the direction between the center of the grating and the center of the CCD detector can be varied between 0$^\circ$ and 15$^\circ$ with an accuracy of 0.001$^\circ$. The CCD chip has 2048\,x\,2048 channels and a pixel size of 13.5\,$\mu$m\,x\,13.5\,$\mu$m over its physical size of 27.6\,mm\,x\,27.6\,mm. This detector size combined with the distances $r_1$ and $r_2$ of the RIXS arm allows to collect RIXS spectra covering an energy range of 35\,eV at a typical incident photon energy of $\sim$530 eV (O K-edge), or a range of 75 eV at an incident energy of $\sim$930 eV (Cu L-edge). The inclination angle $\gamma$ between the surface of the CCD chip and the incoming photon beam can be varied between 15$^\circ$ and 65$^\circ$ with 0.001$^\circ$ precision. By choosing a low inclination angle, the energy resolution of the CCD detector can be improved, because the number of pixels per angle and thus per energy is larger. However, for the smaller inclination angles, a portion of the lower detector channels is shadowed, which should be taken into consideration for experimental planning purposes. By choosing a large inclination angle, the energy range on the detector is increased at the cost of detector resolution.\\
In order to restrict the illuminated area on the active grating, three slits systems (one with vertical aperture, "slit 2v", and two with horizontal apertures, "slit 1h" and "slit 2h" in Fig. \ref{fig:fig7}) are mounted within the grating chamber. These slit systems are also useful to reduce the number of photons that are scattered at the chamber walls and reach the CCD chip indirectly. They are therefore relevant for the suppression of photon background. The positions of these slit systems is shown in Fig. \ref{fig:fig7}. 

The two VLS gratings used in the RIXS spectrometer were produced at HZB. The required substrates were produced by the company Zeiss \citep{Zeiss}. These spherically shaped blanks are made of Silicon and have an outer dimension of 160\,mm\,x\,40\,mm\,x\,40\,mm with an optical surface of 150\,mm\,x\,15 mm. These blanks were used to produce two VLS gratings each having an average line density of 2400\,lines/mm. The parameters of the low and high energy gratings have then been especially optimized for their respective photon energy ranges between 200\,eV - 600\,eV and 400\,eV - 1200\,eV. The complete set of parameters of the two gratings including a comparison between targeted design values and experimentally verified parameters is given in Table \ref{tab:table2}. 

\begin{table*}[t!]
\resizebox{13.5cm}{!}{
\begin{tabular}{c|c|c|c|c}      
\multirow{2}{*}{Parameter} & Low energy grating & Low energy grating & High energy grating & High energy grating  \\
& design & realization & design & realization\\
\hline
 Energy range (eV)     & 200-600      & 200-600      & 400-1200& 400-1200      \\
 Radius (mm)      & 27157 &27071 &41513&41345\\
 Slope error tangential      & \multirow{2}{*}{$<0.1$}      & \multirow{2}{*}{0.111}      & \multirow{2}{*}{$<0.1$}     & \multirow{2}{*}{0.104} \\
 rms, (arcsec) & & & &\\
 Slope error sagittal      & \multirow{2}{*}{$<2.5$}     & \multirow{2}{*}{0.22}      & \multirow{2}{*}{$\leq2.5$}     & \multirow{2}{*}{-} \\
 rms, (arcsec) & & & &\\
Micro roughness      & \multirow{2}{*}{$\leq0.3$}      & \multirow{2}{*}{$<0.19$}      & \multirow{2}{*}{$\leq0.3$}     & \multirow{2}{*}{$<0.28$} \\
 rms, (nm) & & & &\\
 \hline
Linear density N$_0$      & \multirow{2}{*}{2400}      & \multirow{2}{*}{2399.972 $\pm$ 0.007}      & \multirow{2}{*}{2400}     & \multirow{2}{*}{2399.400 $\pm$ 0.005} \\
 (l/mm) & & & &\\
 Linear coefficient      & \multirow{2}{*}{1.0942 x 10$^{-4}$}      & 1.0940 x 10$^{-4}$       & \multirow{2}{*}{1.3058 x 10$^{-4}$}     & 1.3081 x 10$^{-4}$\\
 (mm$^{-2}$) & & $\pm$ 0.0005 x 10$^{-4}$& &$\pm$ 0.0003 x 10$^{-4}$ \\
  Quadratic coefficient      & \multirow{2}{*}{-9.37 x 10$^{-8}$}      & -9.36 x 10$^{-8}$       & \multirow{2}{*}{-8.90 x 10$^{-8}$}     & -8.89 x 10$^{-8}$\\
 (mm$^{-3}$) & & $\pm$ 0.04 x 10$^{-8}$& &$\pm$ 0.02 x 10$^{-8}$ \\
  Cubic coefficient      & \multirow{2}{*}{1.70 x 10$^{-10}$}      & 1.74 x 10$^{-10}$       & \multirow{2}{*}{1.76 x 10$^{-10}$}     & 1.79 x 10$^{-10}$\\
 (mm$^{-4}$) & & $\pm$ 0.07 x 10$^{-10}$& &$\pm$ 0.04 x 10$^{-10}$ \\
  Blaze angle (degrees)     & 2.4      & 2.49 $\pm$ 0.15 (2$\sigma$)      & 1.9     & 2.13 $\pm$ 0.19 (2$\sigma$)\\
  Apex angle (degrees)     & $<174$      & 171.3 $\pm$ 1.6 (2$\sigma$)      & $<170$     & 172.9 $\pm$ 0.9 (2$\sigma$)\\
Micro roughness      & \multirow{2}{*}{$\leq0.6$}      & \multirow{2}{*}{0.27 $\pm$ 0.05}      & \multirow{2}{*}{$\leq0.6$}     & \multirow{2}{*}{0.32 $\pm$ 0.13} \\
 rms, (nm) & & & &\\ 
Au coating thickness      & \multirow{2}{*}{30}      & \multirow{2}{*}{32 $\pm$ 2}      & \multirow{2}{*}{30}     & \multirow{2}{*}{31 $\pm$ 3} \\
 (nm) & & & &\\ 
\end{tabular}
}
\caption{Overview of the two VLS gratings' parameters of PEAXIS. Values in the upper part of the table correspond to the grating substrates provided by Zeiss GmbH \citep{Zeiss}.}
\label{tab:table2}
\end{table*}

\begin{figure}[h]
\centering
\includegraphics[width=7.5cm]{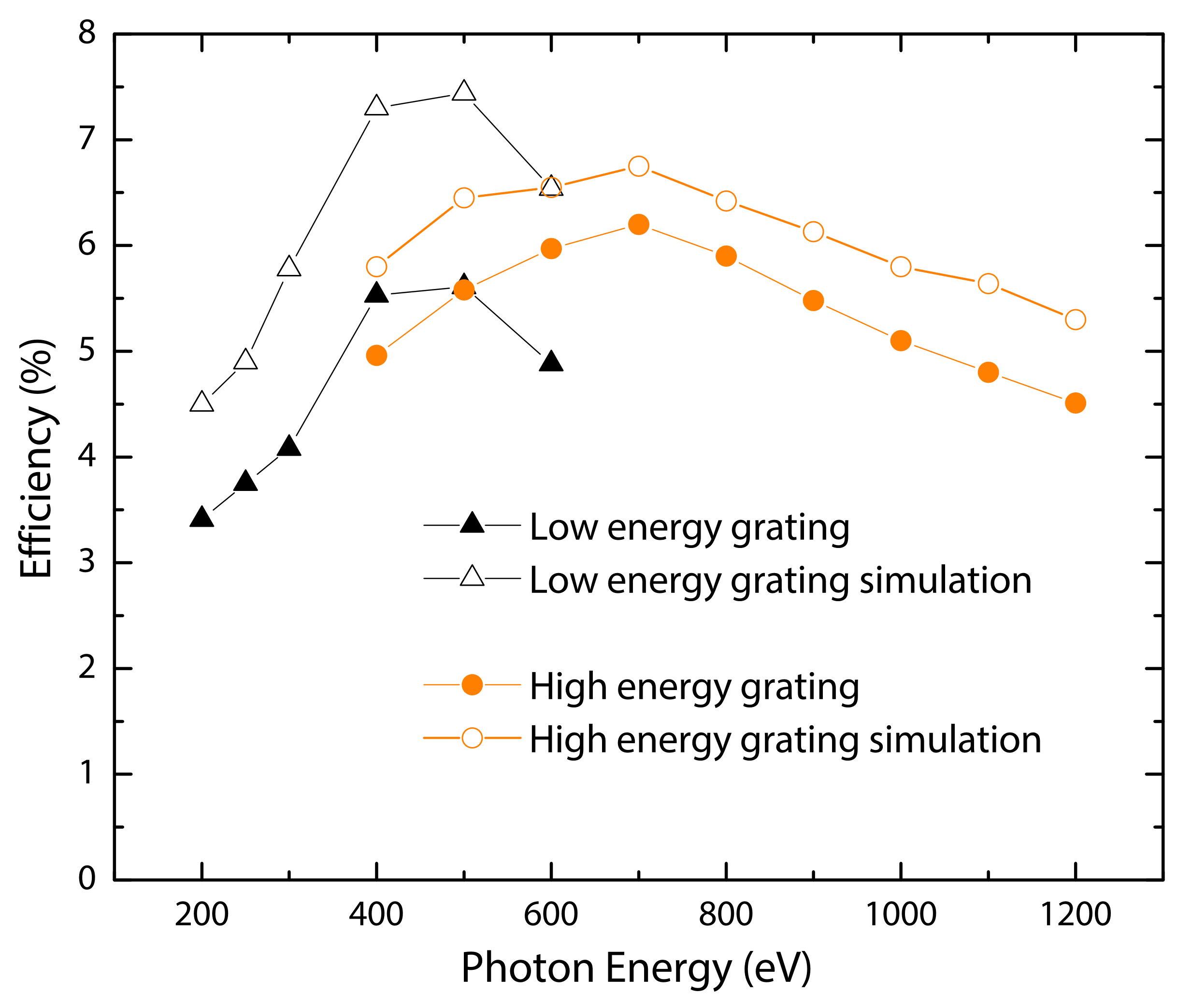}
\caption{Measured (full symbols) and simulated (open symbols) reflection efficiencies of both VLS gratings currently employed at PEAXIS as a function of incident photon energy.}
\label{fig:fig_grating}
\end{figure}

The resulting reflection efficiency of both gratings is shown in Fig. \ref{fig:fig_grating}. It was measured at the reflectometer at the optics beamline at BESSY II \citep{Schaefers2016} and compared with simulations carried out with the ray-tracing software Ray-UI \citep{RayUI} which relies on a code developed by M. Neviere \citep{Neviere1982} to calculate reflection efficiencies of gratings. The overall realized reflection efficiencies for both gratings are comparable to generally achieved values of around 10$\%$, though the absolute values are systematically lower by about 10-20$\%$ than those predicted by the simulation. The reason for this discrepancy is currently being investigated. The general trend of the reflectivity curves as a function of incident photon energy however matches the simulation perfectly.\\

\begin{figure}[b!]
\centering
\includegraphics[width=8cm]{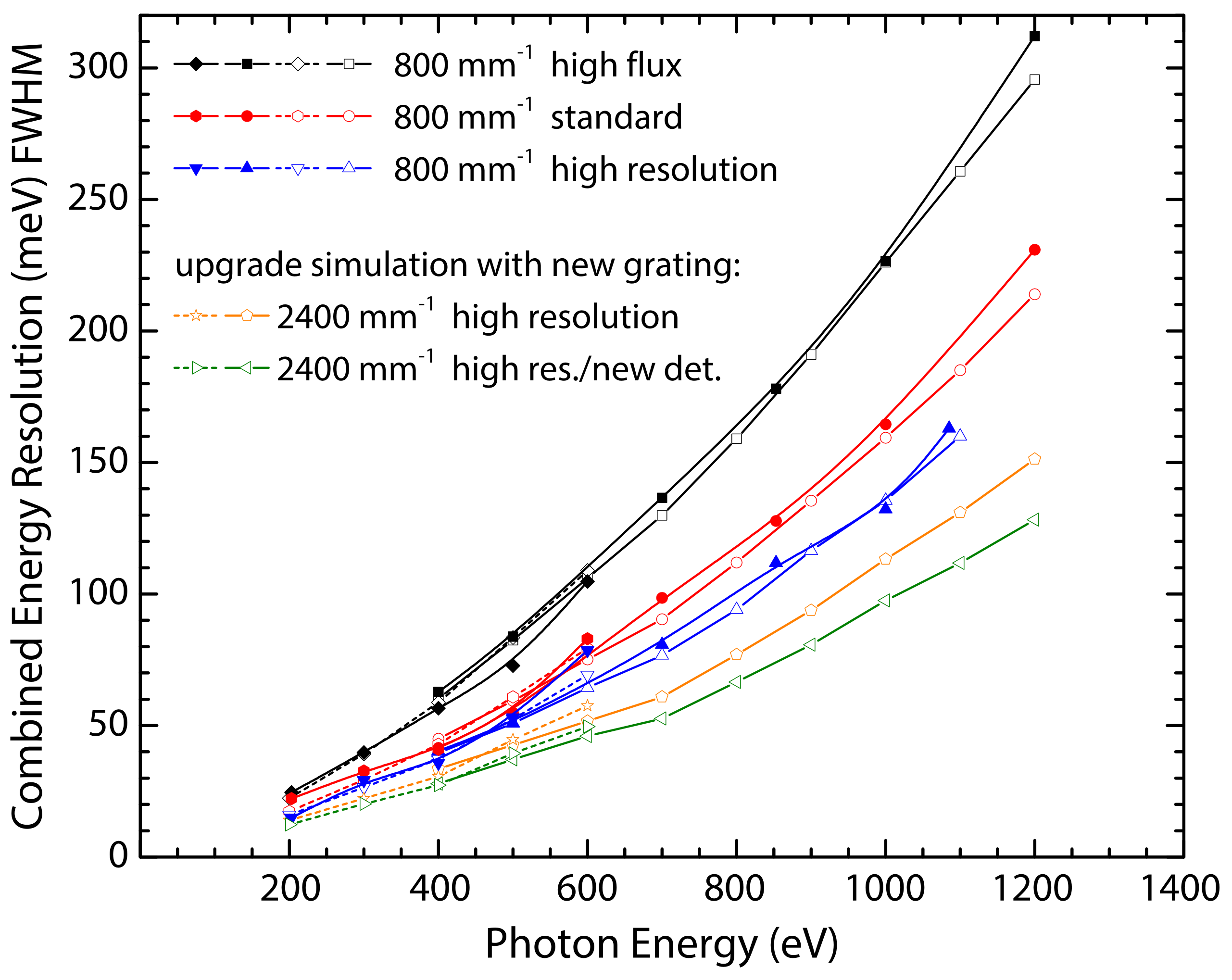}
\caption{Experimentally determined combined energy resolution of PEAXIS for low and high energy gratings (closed symbols) and comparison to simulated resolutions (open symbols). The simulated resolutions for planned upgrades including a new grating with increased line density and/or a new detector with better spatial resolution are also shown.}
\label{fig:fig_combEres}
\end{figure}

The combined energy resolution of the PEAXIS spectrometer over the accessible energy range 200\,-\,1200\,eV for the different operating modes of the beamline have been determined by measuring the RIXS energy width of the elastic scattering at the Ni L$_3$ edge of a polished NiO crystal, positioned in reflection condition. To maximize the intensity and shorten the measurement times, the RIXS spectrometer was operated in backscattering geometry ($2\theta$ = 135$^\circ$). The result of these measurements is shown in Fig. \ref{fig:fig_combEres}. The best resolution obtained on PEAXIS, $\sim20$\,meV, is achieved for an incident photon energy of $\sim200$ eV in the high-resolution mode of operation. For a typical incident energy of $\sim530$ eV (O K-edge), the best energy resolution obtainable for reasonable photon flux is $\sim$50 meV, as shown in Table \ref{tab:table2}. This energy resolution is on a par with that realized at other recently built soft X-ray RIXS instruments at sources with similar brilliance conditions of photon flux.

The beamline and the RIXS spectrometer performance were simulated with the ray-tracing software RAY-UI \citep{RayUI}. The measured resolution is found to be in very good agreement with the simulation results with only slight deviations at the high-energy range limits of each spectrometer grating. In addition to the simulations for the existing monochromator grating with 800\,l/mm, Fig. \ref{fig:fig_combEres} also shows simulation results for a foreseen upgraded monochromator grating with an increased line density of 2400\,l/mm. These simulations were performed for the same parameters as for the 800\,l/mm grating (orange curve in Fig. \ref{fig:fig_combEres}) and for the same parameters in addition to a new detector upgraded to pixel sizes of 5\,$\mu$m in the energy axis (green curve in Fig. \ref{fig:fig_combEres}), respectively. After these planned grating and detector upgrades, a resolving power in excess of 10000 is envisioned for incident photon energies up to 1000\,eV.

\subsection{Photoelectron spectrometer}\label{sec:XPS}

\begin{figure}[h!]
\center
\includegraphics[width=8cm]{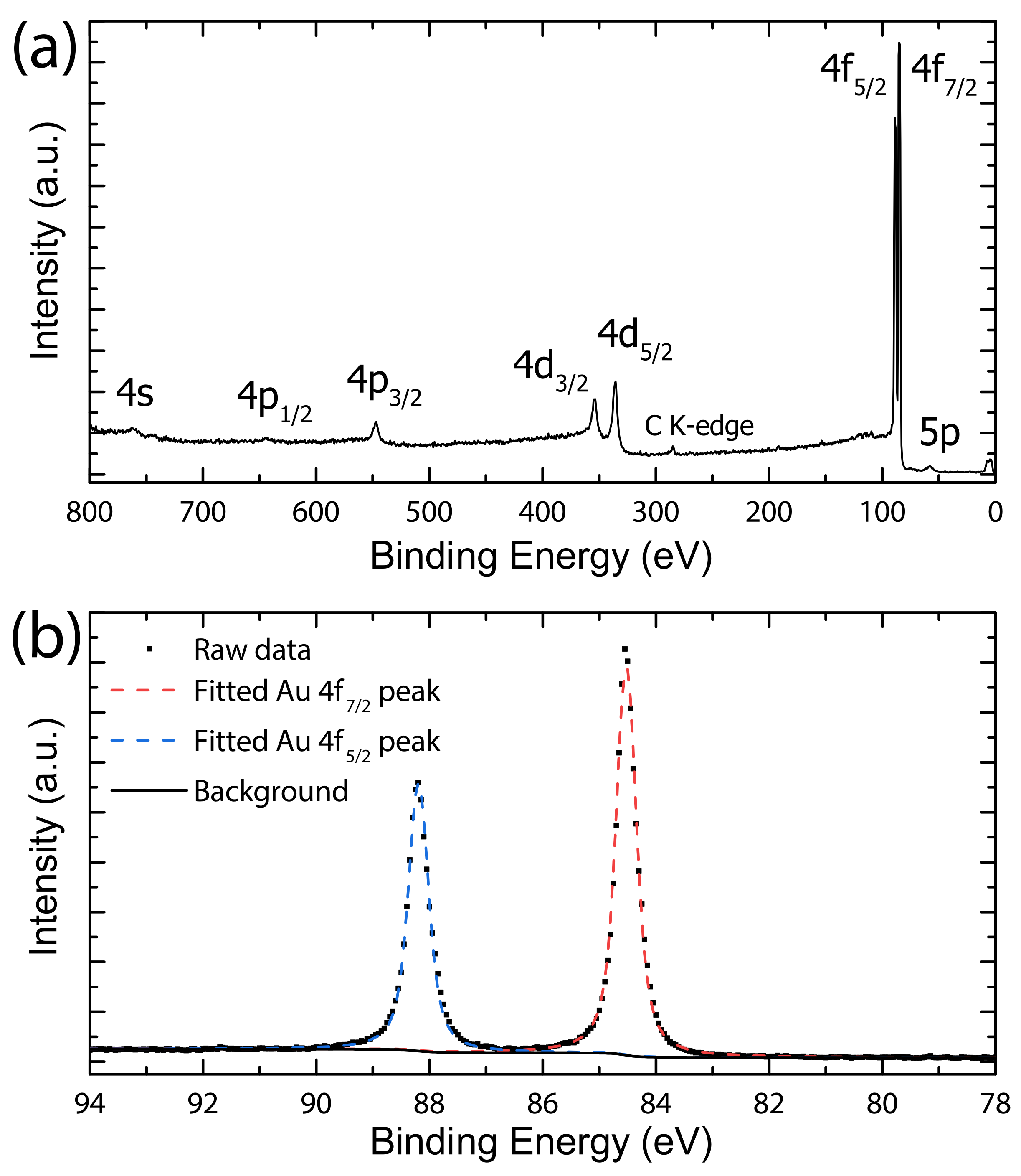}
\caption{Illustration of XPS capabilities of PEAXIS on a 100 nm thick Au film. A survey scan (a) and a fine scan at the Au 4f lines (b) recorded for 1000\,eV incident photon energy and standard beamline mode.}
\label{fig:XPS}
\end{figure}

PEAXIS optionally allows for XPS studies within the same sample chamber. For these studies, photoelectrons which are emitted from the sample after the photon absorption can be collected by an electron analyzer (PHOIBOS 150 EP from SPECS \citep{SPECS}) mounted at the sample chamber (see Figs. \ref{fig:fig5} and \ref{fig:fig7}). In the standard position for XPS measurements with the RIXS arm located at $2\theta=90^\circ$, the XPS detector is positioned at an angle of 55$^\circ$ with respect to the incident X-ray beam in the horizontal and at an angle of 10$^\circ$ in the vertical plane (cf. bottom of Fig. \ref{fig:fig7}). In addition to the possible rotation of the sample about the three manipulator axes $\omega$, $\psi$ and $\chi$, the XPS analyzer itself can be rotated about the beamline axis by an angle $\varphi$ (together with the sample chamber) in order to choose an optimal combination of the angles between the direction of incident photons and sample surface as well as between sample surface and the direction of the emitted electrons. Under normal operating conditions, the sample chamber is connected to the RIXS arm which restricts the angular range of the XPS detector to $-24.8^\circ<\varphi<10^\circ$. This range can be extended to $-28^\circ<\varphi<57^\circ$ upon disconnecting the RIXS arm from the sample chamber (in practice rarely done).\\
The XPS analyzer is equipped with a 2D CCD detector. It is therefore possible to perform angle-dependent measurements. An ultimate energy resolution of $\leq$2\,meV FWHM was confirmed as part of factory acceptance tests by measuring the Xe 5p$_{3/2}$ line at 12.130\,eV binding energy using the He-I excitation line.\\
Being a commercial instrument, the PHOIBOS 150 EP analyzer can be used in different modes of operation. For XPS measurements, the "fixed analyzer transmission" (FAT) mode is typically used, in which the excitation and the pass energies of the analyzer are kept constant whereas the kinetic energy of the electrons is varied. The energy resolution of the analyzer is constant throughout the scan. A typical measurement in this mode to illustrate the XPS capabilities of PEAXIS is shown in Figure \ref{fig:XPS} for a 100 nm Au film (Sigma Aldrich).

The figure shows a survey scan over a wide energy range (a) and a fine scan at the Au 4f lines (b) with the typical energy resolution of 0.43\,eV FWHM of the measurement, which was determined by fitting the data using the CasaXPS software suite \citep{CasaXPS}. For this set of measurements, the beamline was operated in standard mode at an incident photon energy of 1000\,eV, with the PHOIBOS 150 EP set to medium area mode with entrance slit 0.2\,mm\,x\,20\,mm, open exit slit and a pass energy of $E_p=5$\,eV.

In addition to the FAT mode, angle-depended XPS measurements can be performed with three different angular dispersion modes. For occupied-state measurements, the analyzer is typically set to the "constant final state" mode, for which the kinetic energy of the electrons and the pass energy are kept constant while varying the excitation energy. The "constant initial state" mode is correspondingly used for empty-state spectroscopy measurements, in which the excitation energy and the measured kinetic energy of the electrons are varied simultaneously and with equal step size. Finally, Auger electrons are typically collected in the "fixed retarding ratio" mode. In this mode, the excitation energy is kept constant, but the measured kinetic and pass energies are varied to keep the ratio between the retarding potential and the pass energy constant.

\section{Scientific examples}

To illustrate the instrumental capabilities and the performance of PEAXIS, two samples - one solid and one liquid - have been studied by RIXS.

\subsection{Excitations in single crystalline NiO}

RIXS is a coherent photon-in-photon-out process and transitions from initial to final states are not subject to the dipole selection rules as they are with other optical spectroscopies. The technique is particularly suited to probe d-d excitations and charge-transfer excitations in 3d or 4d transition metal complexes. In nickel oxide (NiO), these excitations are intimately related to the six-fold oxygen coordination around each Ni atom. Six oxygen ligands on a regular octahedron cause a splitting of Ni 3d states by symmetry, combined with a Ni-3d + O-2p hybridization.\\

\begin{figure}[h!]
\includegraphics[width=8cm]{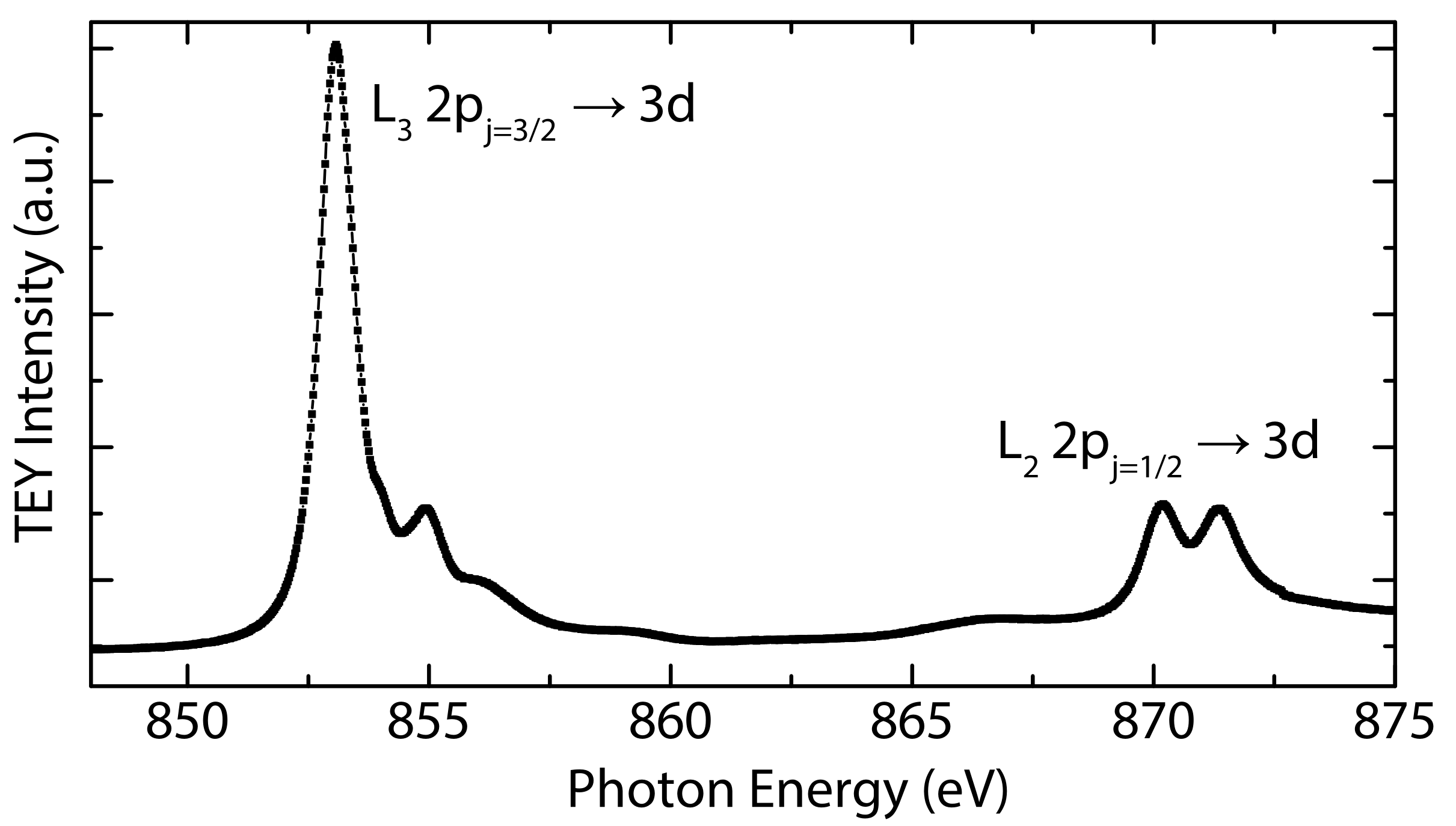}
\caption{TEY spectrum of NiO around the Ni L$_3$ and L$_2$ edges.}
\label{fig:TEY}
\end{figure}

\begin{figure}[b!]
\includegraphics[width=8cm]{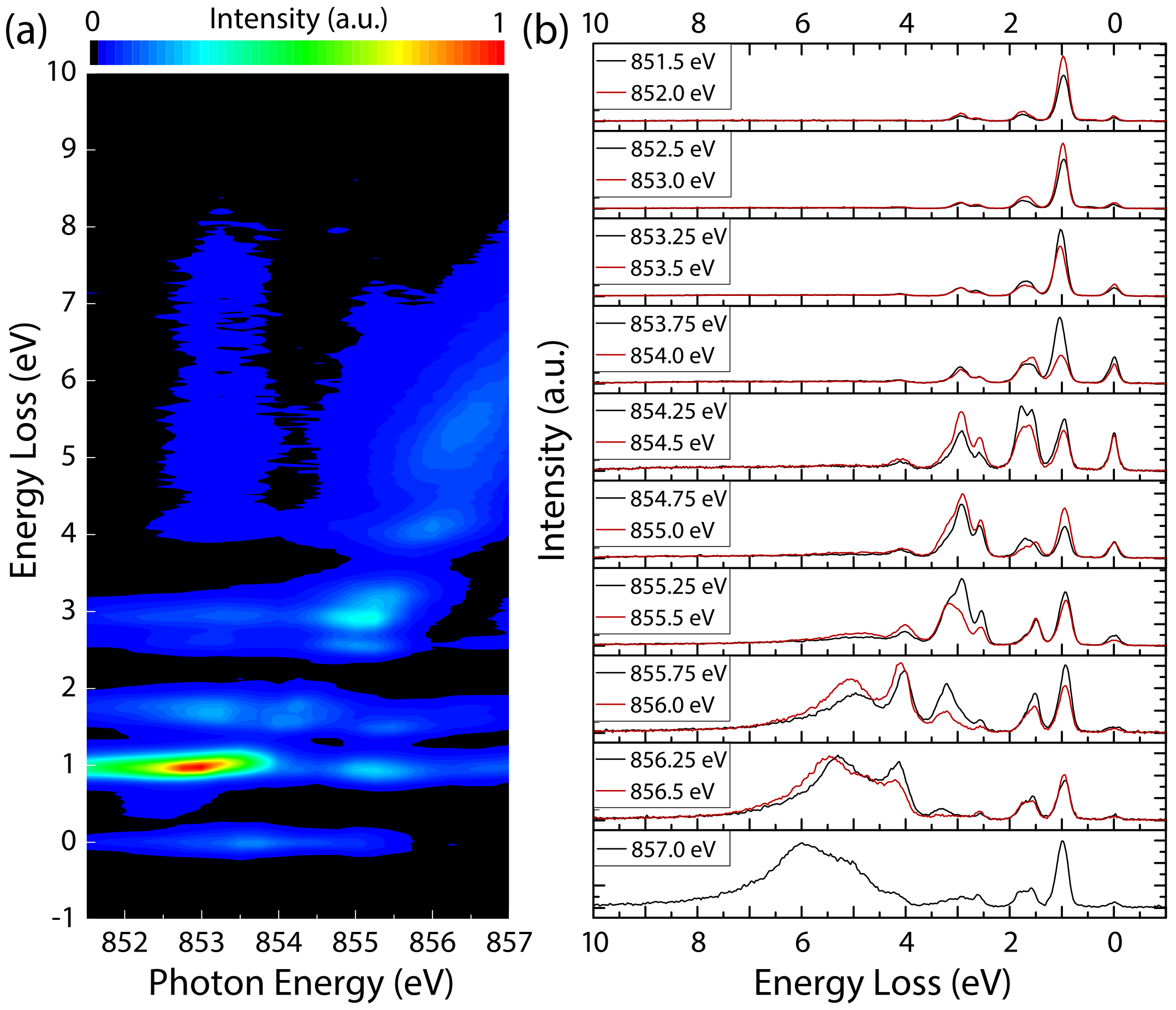}
\caption{Two-dimensional RIXS map (a) of NiO displaying regions with high intensity corresponding to the d-d excitations at constant energy transfers and charge transfer excitations for energy transfers continuously increasing with increasing incident photon energy, and the raw RIXS spectra collected (b) to create the RIXS map.}
\label{fig:fig_NiOmap}
\end{figure}

To probe the electronic Ni 3d states in NiO, incident photon energies were chosen close to the Ni L$_3$ resonance around 853\,eV. For that purpose, total electron yield (TEY) measurements were first performed to accurately determine the L$_3$ resonance energy as shown in Fig. \ref{fig:TEY}. As the emitted RIXS spectrum depends sensitively on the incident photon energy (see Fig. \ref{fig:fig_NiOmap}(b)), the precise resonance energy for the L$_3$ (and L$_2$, or any absorption edge of interest) excitation should be determined by TEY upon X-ray absorption before the collection of RIXS spectra. Figure \ref{fig:fig_NiOmap} shows experimental data measured at PEAXIS which are in good agreement with experimental and calculated data by Betto \textit{et al.} \citep{Betto2017} and with earlier data by Ghiringhelli \textit{et al.} \citep{Ghiringhelli2005}.

The RIXS spectra shown in Fig. \ref{fig:fig_NiOmap}(b) were collected in the standard mode of the beamline yielding an energy resolution of 136\,meV with incident photon energies from 851.5 eV well below the L$_3$ absorption edge to 857.0 eV  well beyond the L$_3$ edge. To avoid strong elastic scattering from the sample, the scattering angle of the RIXS spectrometer was set to 110$^\circ$ close to the linear polarization direction of the incident X-ray beam. In order to even further suppress the elastic scattering, the sample was rotated by an angle of 4$^\circ$ away from the elastic reflection conditions at $\theta=59^\circ$. Each spectrum was measured for 15 min.

For incident photon energies below the L$_3$ resonance of 853\,eV, there is very little variation in the measured spectra. Above the resonance, new features become more pronounced with increasing energy. One part of these spectra is related to the d-d excitations (up to about -4\,eV), which can be described within the crystal field model of NiO \citep{Uldry2012}. The second, dispersing part at higher energy transfers is related to the charge transfer excitations and can be described by the Anderson impurity model \citep{Ghiringhelli2005,Matsubara2005,Magnuson2002}.\\ 
The main peak at -1 eV (which is present in all measured spectra) can be interpreted as corresponding to the crystal field splitting parameter $10Dq$, which is given as the energy splitting between the singly occupied e$_g$ ($x^2 - y^2$, $z^2$) and t$_{2g}$ ($xy$, $yz$, $zx$) orbitals in the crystal field model. All peaks between -1\,eV and -4\,eV are caused by a redistribution of the 3d electrons among the available orbitals during the scattering process (localized at the Ni ion) from an initial 3d$^8$ ground state to a new configuration 3d8$^*$ at higher energy with respect to the ground state, effectively leading to an excited d-state, thus called d-d excitation. This energy difference is detected in the RIXS measurement as an energy loss of the scattering process. If the incident photon energy is high enough to populate also the configuration with a ligand hole (3d$^{\underline{9}}$\underline{L}), then a charge transfer excitation with a larger energy loss can occur. The energy difference between these two configurations  (3d$^{\underline{9}}$\underline{L}, 3d$^8$) is called charge transfer energy $\Delta$ =  E(3d$^{\underline{9}}$\underline{L}) - E(3d$^8$). A more detailed analysis can be found in Ghiringhelli \textit{et al.} \citep{Ghiringhelli2005} and references therein. \\
To get a comprehensive overview of all features and their intensity as a function of the incident photon energy, the individual RIXS spectra can be combined into a two-dimensional RIXS map, shown in Fig. \ref{fig:fig_NiOmap}(a).

\begin{figure}
\includegraphics[width=7.5cm]{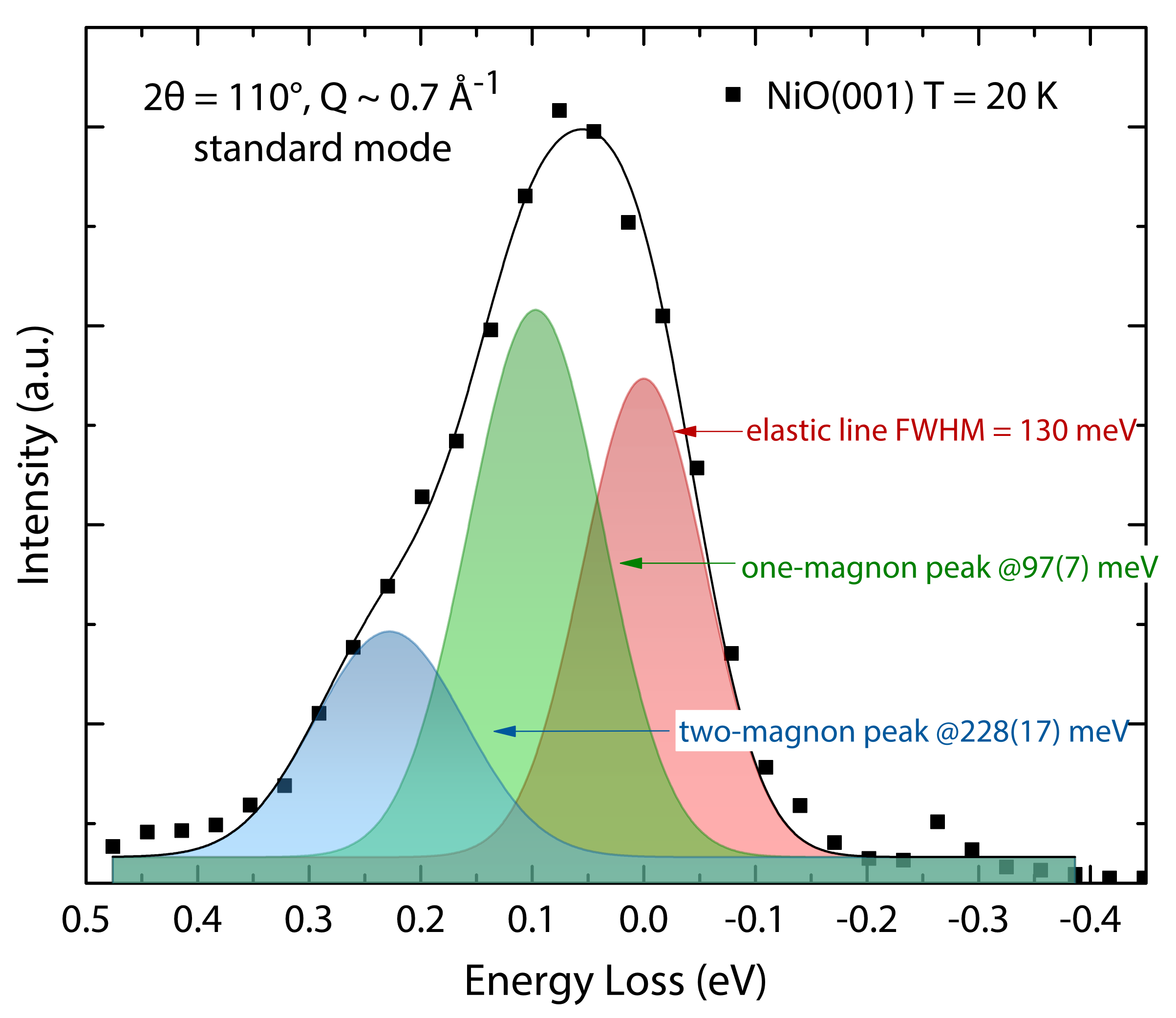}
\caption{One- and two-magnon excitation in NiO close to the elastic signal.}
\label{fig:fig_NiOmagnon}
\end{figure}

In addition to probing electronic excitations, RIXS allows to probe magnetic excitations in single crystalline samples \citep{Betto2017,Ament2009PRL,deGroot1998}. The mechanism by which the magnetic excitations are created depends however on the particular RIXS process. Spin and orbital degrees of freedom couple if the intermediate state for example at an L-edge involves a core-hole with strong spin-orbit coupling. In a direct process, the angular momentum of photons can then be transferred to the spins to create a magnetic excitation (spin flip excitation) \citep{deGroot1998,AmentPhD}. If the RIXS process involves a core-hole without spin-orbit coupling, such as for example at a K-edge, the transient intermediate state is a local, magnetic impurity that perturbs the system and causes a magnetic excitation \citep{AmentPhD}. This indirect RIXS channel exists in principle also for spin-orbit coupled core-holes but has been shown to contribute only in higher order to the spectral weight.\\ 
Probing low-energy magnetic excitations is challenging, as their energies are typically one or two orders of magnitude smaller than those of electronic d-d excitations. The success of the measurements therefore depends on the energy resolution the spectrometer can provide at the relevant resonance edges. Figure \ref{fig:fig_NiOmagnon} shows a one-magnon and a two-magnon excitation in NiO(001) measured with an energy resolution of 130 \,meV in the standard mode of operation at a sample temperature of T $=$ 20\,K.\\
This RIXS measurement was performed at the Ni L$_3$ edge to maximize the spectral weight of the one-magnon excitation. As previously discussed, a scattering angle of $2\theta = 110^\circ$ and an angle $\theta=59^\circ$ was selected to reduce the elastic scattering from the sample. At this angular setting, the wave vector transfer $Q$ is $\sim0.7$ \AA$^{-1}$, which corresponds to the Brillouin zone boundary along the [110] direction and which thus allows to observe the maximum of the magnon dispersion at $\sim110$ meV \citep{Hutchings1972}. The beamline operated in the high-resolution mode with exit slit size 5 $\mu$m and c$_{\rm{ff}}$=5. The data accumulation time per spectrum was 60 min. The one-magnon peak was fitted at 97 meV $\pm$ 7 meV with a FWHM of 141 meV $\pm$ 6 meV and the two-magnon peak at 228 meV $\pm$ 17 meV with FWHM of 160 meV $\pm$ 30 meV. 

Our results demonstrate the feasibility to study magnetic excitations with RIXS at PEAXIS. They are in good agreement with recent results from Betto \textit{et al.} \citep{Betto2017} who  performed magnon dispersion measurements at the ID32 beamline at the ESRF with a resolving power that was by a factor of 3 better than that presently achievable at PEAXIS, corresponding to an energy resolution of 40\,meV at that beamline. Betto \textit{et al.} observe the one-magnon peak at around 100 meV for $Q = 0.7$ \AA$^{-1}$ with a FWHM width of 50 - 60\,meV and the two-magnon peak at 190 meV with a FWHM of $\sim$150\,meV. The peak positions extracted from our measurements agree well with these latter results.

\subsection{Vibrational excitations in liquid acetone}

\begin{figure}
\centering
\includegraphics[width=7.5cm]{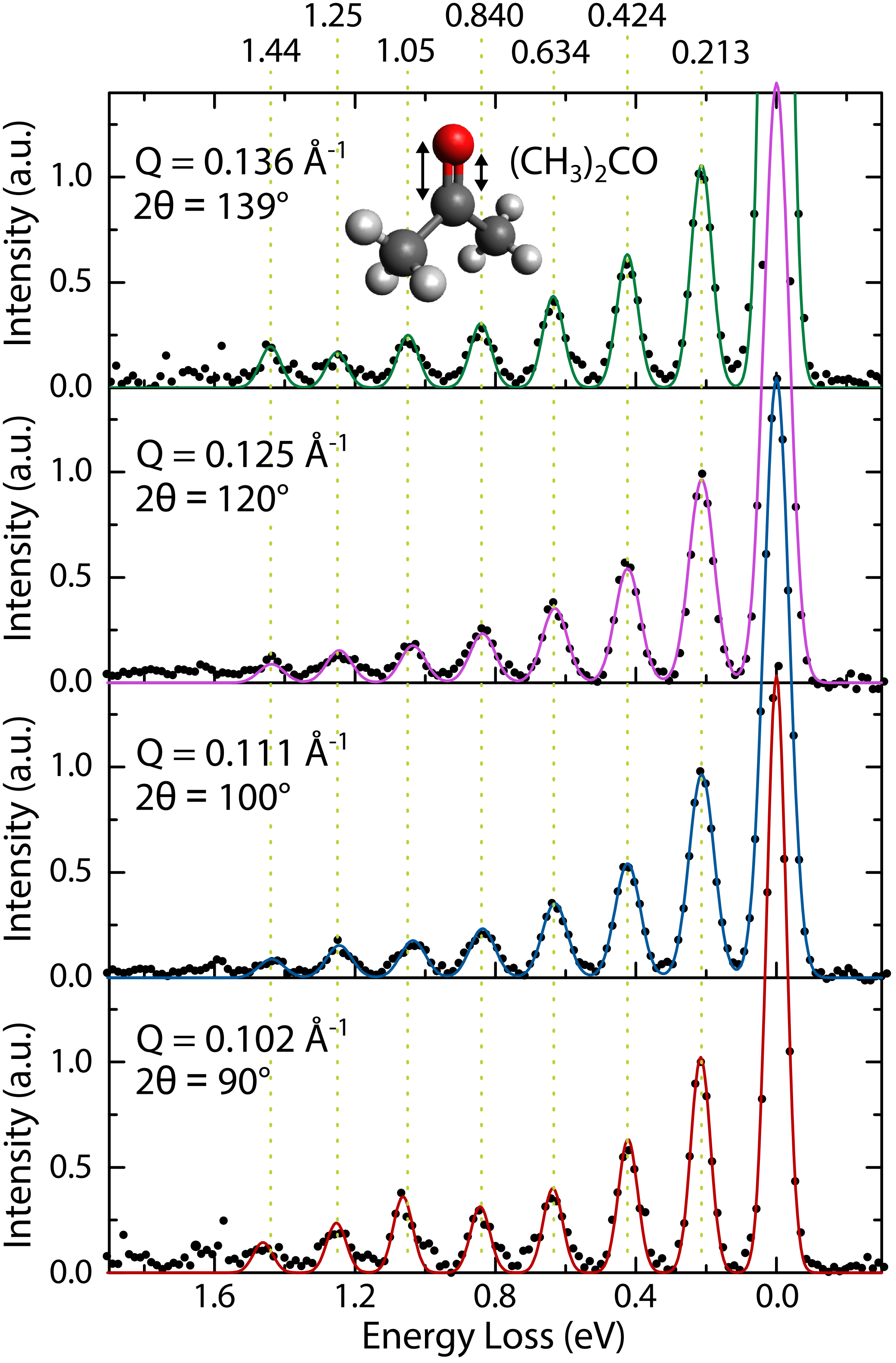}
\caption{The vibrational excitations in liquid acetone were measured at the Oxygen K-edge of 531\,eV with an energy resolution of 63 meV. The intensity is normalized to the first vibrational peak.}
\label{fig:fig_acetone}
\end{figure}

Using RIXS, vibrational modes in individual molecules in the liquid state \citep{Rubensson2013,Schreck2016} and crystals \citep{Ament2011EPL,Lee2013} can be investigated. This particular scattering channel is a direct consequence of the coupling between the electronic charge distribution in the intermediate state and the nuclear dynamics \citep{Ament2011EPL}, based on the idea that the locally perturbed charge distribution in the transient intermediate state displaces the neighboring ions from their equilibrium positions and therefore causes molecular or lattice vibrations \citep{Veenendaal2015}.\\
Here, RIXS data from liquid acetone are shown in Fig. \ref{fig:fig_acetone}. The presented spectra illustrate the capability of PEAXIS to study vibrational excitations in soft or liquid matter samples. For these measurements, a static fluid cell with a 75\,nm thin entrance window made of Si$_3$N$_4$ of 0.5\,mm x 0.5\, mm size was used to measure liquid acetone (volume of 5 $\mu$l) in the ultra high vacuum of the sample chamber. The measurements were done in the standard mode of operation of the beamline to have sufficient flux to compensate for the X-ray absorption in the entrance window of the cell. The incident energy was set to 531 eV (Oxygen K-edge) to excite a 1s electron into a $\pi^*$ anti-bonding orbital of acetone \citep{Schreck2016}. The resulting energy resolution at that incident photon energy was $\sim$\,63\,meV, which is sufficient to resolve the C=O stretching mode of liquid acetone at a fundamental energy of $\sim210$\,meV \citep{Schreck2016}. To investigate whether the wave vector $Q$ has an influence on the vibrational progression, RIXS spectra at four different scattering angles ($2\theta$ = 90$^\circ$, 100$^\circ$, 120$^\circ$ and 139$^\circ$) have been recorded. 
With the current signal-to-noise ratio of PEAXIS, seven peaks at the fundamental frequency and its overtones can be confidently identified (see Fig. \ref{fig:fig_acetone}). These could be satisfactorily described by resolution-limited Gaussian lineshapes, resulting in the following peak positions: 213$\pm$2 meV, 424$\pm$2 meV, 634$\pm$3 meV, 840$\pm$3 meV, 1048$\pm$12 meV, 1248$\pm$6 meV and 1445$\pm$12 meV, in excellent agreement with previously reported results by Schreck \textit{et al.}. The four spectra do not show any dependence on scattering angle and thus no noticeable wavevector dependence in the small $Q$-region probed by the RIXS measurements in this setup. 

\section{Conclusion and outlook}

PEAXIS is a versatile X-ray spectrometer offering $Q$-dependent RIXS measurements on solid and liquid samples over a wide temperature range. The performance of the beamline and the RIXS spectrometer were demonstrated by measurements on N$_2$ and Ne gas as well as by measurements of typical excitations: Charge transfer excitations and d-d excitations were investigated at the L$_3$ edge of Ni within a high-quality NiO crystal. The same crystal was used to measure single and double magnon excitations. To demonstrate the capability to measure liquid samples, the vibrational excitations in liquid acetone were also presented. The results from these measurements clearly show that PEAXIS is very well suited for the study electronic excitations and, limited by the energy resolution, also for the study of quasi-particle excitations. On PEAXIS, magnetic and phonon excitations at $\sim$100 meV can be resolved at typical transition metal L-edges energies. 
To improve the energy resolution of the beamline, its monochromator will be upgraded by an additional grating of higher line density (2400\,lines/mm, cf. Fig. \ref{fig:fig_combEres}). After this upgrade, the energy resolution of the beamline and the RIXS spectrometer will be matched. Furthermore, a replacement of the undulator at the beamline is foreseen. The new undulator will provide a higher photon flux, especially at photon energies above 1200\,eV and, in addition, an adjustable photon beam polarization to tailor to the experimental requirements. An upgrade of the RIXS CCD detector to a model with smaller pixel size will further improve the energy resolution, targeting a resolving power of $\geq$\,10000 at 1000\,eV.




\section*{Acknowledgments}
We would like to thank Peter Guttmann and Matthias Mast for their great work building up and commissioning the main part of the U41-PEAXIS beamline. The help of the teams from the HZB Department Optics and Beamlines, the Department Precision Gratings, the Department Nanometre Optics and Technology and the Department Scientific-Technical Infrastructure is gratefully acknowledged. The project was funded in part by the German BMBF under F\"orderkennzeichen 05K13KE4.

\bibliography{PEAXIS_instrumentpaper}






\end{document}